\documentclass[epj]{svjour}
\usepackage{graphicx}
\begin{document}
\title{Exact curvilinear diffusion coefficients 
in the repton model.}
\author{Arnaud Buhot}
\institute{UMR 5819 (UJF, CNRS, CEA) SPrAM, D\'epartement
de Recherche Fondamentale sur la Mati\`ere Condens\'ee,
CEA Grenoble, 17 rue des Martyrs, 38054 Grenoble cedex 9, 
France. \email{abuhot@cea.fr}}

\abstract{
The Rubinstein-Duke or repton model is one of the simplest 
lattice model of reptation for the diffusion of a polymer 
in a gel or a melt. Recently, a slightly modified model 
with hardcore interactions between the reptons has been 
introduced. The curvilinear diffusion coefficients of both 
models are exactly determined for all chain lengths. The 
case of periodic boundary conditions is also considered.
}
\date{\today}

\maketitle

\section{Introduction}

The reptation of a polymer in an entangled melt was 
studied long ago by De Gennes~\cite{DeGennes} who 
predicted the polymer length dependences of the
curvilinear and self diffusion coefficients as well
as the viscosity and the relaxation time. Later, 
Rubinstein introduced a lattice model for the polymer 
motion incorporating most of De Gennes' ideas of 
reptation~\cite{Rubinstein}. The so called repton 
model was then generalized by Duke to take into 
account the case of charged polymers during gel 
electrophoresis allowing the determination of the 
drift velocity~\cite{Duke}. This repton model seems 
particularly well adapted for DNA gel electrophoresis 
when the pore size is comparable to the persistence 
length~\cite{Viovy,Slater}.

The theoretical prediction of the viscosity dependence 
with the polymer length is in apparent conflict with the 
experimental observations~\cite{Viovy,Milner,Lodge,Wang}. 
This discrepancy is also observed in numerical simulations 
of the repton model~\cite{Rubinstein}. Furthermore, a lot 
of interest in the calculation of the self-diffusion in 
the repton model focussed  on the long polymer limit. 
The next to leading order term of the self-diffusion was 
long debated due to discrepancy between analytical and 
numerical results~\cite{Widom,Barkema,Prahofer,Newman,Carlon}.

At the same time, the diffusion of a polymer chain in 
small channels attracts an increasing interest since 
it applies to a great range of experimental situations.  
Brochard and De Gennes~\cite{Brochard} considered the 
case of a flexible polymer in a channel large compared 
to the monomer size but small compared to the polymer 
length whereas Odijk~\cite{Odijk} studied the case of
stiff polymers with a persistence length larger than 
the channel width. The recent experimental access to 
nanometer scale channels allows to study the crossover 
behaviour between both regimes~\cite{Tegenfeldt,Reisner}. 
The transport of long flexible polymer chains through 
Carbon nanotubes bring the interest to channels of width 
comparable to the monomer size allowing the determination 
of the curvilinear diffusion coefficients~\cite{Wei}. 
Furthermore, the case of polymer diffusion in porous 
media with nanometer scale holes has been recently
studied~\cite{Terranova} with a slightly modified repton 
model~\cite{Guidoni}. This model presents identical 
dynamical rules for the curvilinear motion of the 
chain than the repton model and motivated the present 
interest on the analytical determination of the 
curvilinear diffusion coefficients as function of 
the chain length. 

The paper is organized as follows. In section 2 the 
repton model is presented with the slight modifications
introduced by Guidoni et al.~\cite{Guidoni}. In section 
3, the main result of the paper, the exact calculation 
of the curvilinear diffusion coefficients, is presented
and compared with previous results. Section 4 focuses 
on the repton model with periodic boundary conditions.
Finally, in section 5, we discuss the next to leading 
order corrections of the exact curvilinear diffusion 
coefficients and we give some conclusions. 

\section{Description of the model}

In the Rubinstein or repton model~\cite{Rubinstein}, 
the polymer chain contains $N$ beads or reptons. 
A configuration ${\cal C}$ is characterized by $N-1$ 
variables $\tau_i$ corresponding to the existence 
($\tau_i = 0$) or not ($\tau_i = 1$) of a stored 
length between the two reptons $i$ and $i+1$ 
along the chain. The polymer diffusion in a melt 
or a gel is obtained by the motion of stored lengths 
along the chain. Duke~\cite{Duke} generalized the model 
incorporating the diffusion along a spatial direction 
subject to an electric field in order to model the 
electrophoresis of a polymer in a gel. The variables 
$\tau_i$ are then replaced by $N-1$ variables $\sigma_i$ 
with $\sigma_i = 0$ corresponding to the existence of 
a stored length and $\sigma_i = \pm 1$ to the direction 
of the chain along the field between the reptons $i$ 
and $i+1$ in the absence of stored length. In the simple 
Rubinstein model, the number of configurations is 
$2^{N-1}$ whereas it is increased to $3^{N-1}$ in 
the Duke model. We are interested in the curvilinear 
diffusion coefficient of a neutral polymer or in 
absence of an electric field. In those cases, the 
direction of the chain between two reptons is not 
relevant. Thus, we restrict ourself to the Rubinstein 
model in the following. This reduces considerably the 
number of configurations to consider without limitations
on the generality of the results. 

\begin{figure}
\begin{center}
\includegraphics[width=7cm]{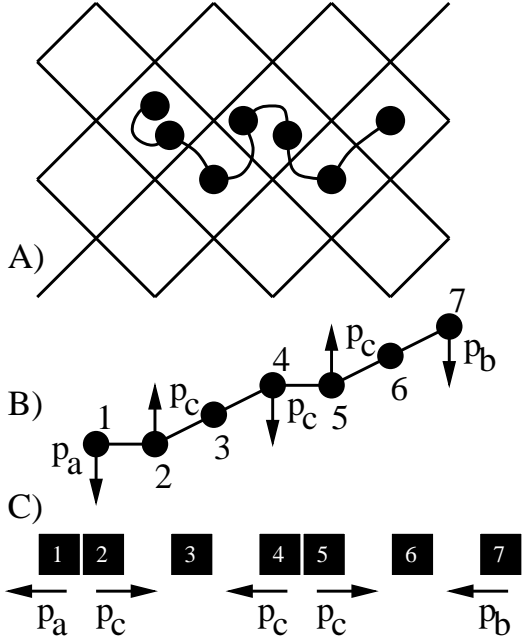}
\caption{\label{moves}
A) A two dimensional representation of a polymer 
chain with $7$ reptons and the underlaying cell 
structure in the repton model.
B) The curvilinear representation of the same chain 
with horizontal lines linking two reptons in the 
same cell and with inclined lines for two reptons 
in different cells. The arrows represent the possible 
moves of the reptons with their respective rates 
($p_a, p_b$ or $p_c$). 
C) The representation of a polymer chain in the 
model with hardcore reptons. The existence of a 
gap between two reptons represents a hole. 
The possible moves of the reptons and their 
respective rates are represented by arrows.
The identical curvilinear dynamics of both models 
is evident comparing the rates in B) and C).}
\end{center}
\end{figure}

In the Rubinstein model, three different moves 
exist (see Fig.\ref{moves}): a) the motion of 
an end repton outside its cell when this cell 
is occupied by the neighbor repton, b) the motion 
of an end repton from its own cell to the cell 
of its neighbor if both cells differ, and c) the 
motion of an internal repton. Cell occupancies 
need to be consistent with this move which means 
that the repton leaves a cell occupied by a 
neighbor and reaches the cell of the other 
neighbor. The moves a) and b) correspond to the 
exit and entrance of a stored length in the chain 
whereas c) is the internal motion of a stored 
length. The respective rates of the moves are 
$p_a$, $p_b$ and $p_c$. In general, those rates 
are proportional to the number of accessible cells 
for the moving repton. Thus, in $d$ dimensions and 
with a cubic lattice, $p_b = p_c$ and $p_a = 2 d p_b$. 
However, the existence of larger or smaller monomers 
at the two ends of the chain may either reduce or 
increase the ratio $p_b/p_c$. Similarly, in case of 
a polymer embedded in a gel, the gel structure in which 
the polymer diffuses may affect the ratio $p_a/p_b$. 
The later ratio is responsible as we shall see to the 
equilibrium properties and especially the curvilinear 
length of the polymer chain. The former ratio only 
affects the dynamics of the polymer.

Guidoni et al.~\cite{Guidoni} introduced recently 
a slightly modified model with hardcore reptons in 
one dimension. The cells may only be occupied by a 
single repton and two neighbor reptons are either 
in neighbor cells ($\tau_i =0$) or in next neighbor 
cells ($\tau_i = 1$). The description of the chain 
in terms of the variables $\tau_i$ is thus identical 
to the repton model. $\tau_i = 1$ corresponds to a 
hole between reptons and $\tau_i = 0$ to the existence 
of a stored length. From the repton model, only the 
length of the chain is slightly modified to take into 
account the repton size. The polymer diffuses thanks 
to the motion of stored lengths or holes. Furthermore, 
as seen on Fig.~\ref{moves}, the dynamical properties 
of the curvilinear motion are identical for both models. 
As a consequence, the curvilinear diffusion coefficients 
are equal. 

\section{Exact curvilinear diffusion coefficients}

The curvilinear positions of the different reptons 
along the chain are defined by $s_i = s_1 + a 
\sum_{k < i} \tau_k$ with $s_1$ the position of the 
first repton and $a \tau_i$ the length between the 
two reptons $i$ and $i+1$. In case of hardcore reptons, 
an extra term $(i-1) a$ should be added to take into 
account the repton lengths. It is also interesting 
to define $s_c = \sum_i s_i/N$ the curvilinear position 
of the center of mass as well as $s_m = (s_1 + s_N)/2$ 
the middle position of the polymer chain. This middle 
position differs from the center of mass one for a 
particular configuration ${\cal C}$: $\delta s = s_c - 
s_m = a \sum_i (N-2 \, i) \tau_i /2 N$. This difference 
leads to an internal force that drives the polymer
as will be shown later. 

Due to the $N-1$ variables $\tau_i$ and their two 
possible values, there exist $2^{N-1}$ internal 
configurations ${\cal C}=\left\{\tau_i \right\}_{i 
= 1,\dots,N-1}$ for the polymer chain. We define the 
probability $P({\cal C},s,t)$ for the configuration 
${\cal C}$ to have a curvilinear center of mass position 
$s_c = s$ at time $t$. Due to the correlations between 
the internal configurations of the chain along the time, 
the  evolution of the global variable $s$ alone is not 
sufficient to determine the curvilinear diffusion 
coefficient. The initial condition $P({\cal C},s,t = 0) 
= P_{eq}({\cal C}) \delta (s)$ is assumed with $\delta 
(s)$ the usual delta function and $P_{eq}({\cal C})$ the 
equilibrium probability of the configuration ${\cal C}$. 
From this initial condition, the averaged curvilinear 
position over the ensemble of configurations remains 
zero at all times. Furthermore, the averaged squared 
curvilinear position increases linearly with time 
proportionally to twice the curvilinear diffusion 
coefficient $D_c (N)$ of the polymer chain with 
$N$ reptons. Note that we may expect the average 
position for a single configuration ${\cal C}$ to 
shift a finite value from zero due to the initial 
difference $\delta s$ between the middle and the 
center of mass of the chain. The knowledge of this 
shift in position is necessary to determine the 
curvilinear diffusion coefficients. In fact, it 
reflects the correlations inside the chain between 
the configurations~\cite{Guidoni}. 

\begin{table}
\begin{center}
\begin{tabular}{|c|c|}
\hline
$i$ & $w({\cal C}_i \rightarrow {\cal C})$ \\
\hline
$0$ & $p_{a} \tau_1 + p_{b} (1-\tau_1)$\\
\hline
$1$ to $N-2$ & $p_{c} (\tau_i + \tau_{i+1} 
- 2 \tau_i \tau_{i+1})$ \\
\hline 
$N-1$ & $p_a \tau_{N-1} + p_b (1-\tau_{N-1})$ \\
\hline
\hline
$i$ & $w({\cal C} \rightarrow {\cal C}_i)$ \\
\hline
$0$ & $p_{a} (1-\tau_1) + p_{b} \tau_1$\\
\hline
$1$ to $N-2$ & $p_{c} (\tau_i + 
\tau_{i+1} - 2 \tau_i \tau_{i+1})$ \\
\hline 
$N-1$ & $p_a (1-\tau_{N-1}) + p_b \tau_{N-1} $ \\
\hline
\end{tabular}
\end{center}
\caption{\label{duke}Rates $w({\cal C}_i \rightarrow 
{\cal C})$ and $w({\cal C} \rightarrow {\cal C}_i)$ 
in the repton model. The variables $\tau_i$ corresponds
to the configuration ${\cal C}$.}
\end{table}

In order to define the dynamical rules, we introduce 
the configurations ${\cal C}_i$ for $i = 0$ to $N-1$ 
which only differ from the configuration ${\cal C}$ 
by the fact that both variables $\tau_i$ and 
$\tau_{i+1}$ have changed to $1 - \tau_i$ and 
$1 - \tau_{i+1}$. The changes from ${\cal C}$ to 
${\cal C}_i$ are the only allowed moves in the repton 
model and it corresponds to a stored length passing 
through the repton $i+1$ (see Fig.\ref{moves}). 
In order for the move to exist the variables $\tau_i$ 
and $\tau_{i+1}$ must differ. With those constraints 
in mind, the evolution equation for the probability 
$P({\cal C},s,t)$ is:
\begin{eqnarray}
\label{EqDif}
\frac{\partial}{\partial t} P({\cal C},s,t) & = & -
\sum_{i=0}^{N-1} w({\cal C} \rightarrow {\cal C}_i) 
P({\cal C},s,t) \\
\nonumber & & + \sum_{i=0}^{N-1} w({\cal C}_i 
\rightarrow {\cal C}) P({\cal C}_i, s + \Delta s 
({\cal C}),t) 
\end{eqnarray}
where the rates $w$ are given in Table~\ref{duke}
for open boundary conditions and $\Delta s ({\cal C}) 
= (\tau_{i+1}-\tau_i) a/N$ corresponds to the
curvilinear motion of the center of mass. The 
variables $\tau_0 = 1-\tau_1$ and $\tau_N = 1 -
\tau_{N-1}$ have been introduced for consistency. 
The rate $p_a$ corresponds to the terminal repton 
exploring a new cell, the rate $p_b$ to the entrance 
of the terminal repton into the cell of its neighbor
and the rate $p_c$ to the motion of stored length 
inside the chain (see Fig.\ref{moves}). Those rates 
are identical to those of the model of Guidoni et 
al.~\cite{Guidoni}.  

The determination of the curvilinear diffusion 
coefficients depends on three steps: the determination 
of i) the equilibrium probability $P_{eq}({\cal C})$, 
ii) the long time limit of the averaged curvilinear 
position of a configuration ${\cal C}$ and iii) the 
long time limit of the derivative of the averaged 
squared curvilinear position (in the last case the 
average stands for the position as well as the 
configuration averages). The different moments of 
the curvilinear positions of the center of mass 
are defined by:
\begin{equation} 
\int s^k P({\cal C},s,t) ds \equiv \langle s^k 
\rangle ({\cal C},t) P({\cal C},t)
\end{equation}
where $P({\cal C},t)$ is the probability to find the 
configuration $\cal C$ at time $t$. The time evolution 
for this probability is given by Eq.~\ref{EqDif} where 
$P({\cal C},s,t)$ is replaced by $P({\cal C},t)$. 
From our choice of the initial condition, $P({\cal C},t) 
\equiv P_{eq}({\cal C})$ at all times and satisfies 
Eq.~\ref{EqDif} with $\partial P/\partial t = 0$. 
The equilibrium probability $P_{eq}({\cal C})$ is the 
product of an identical probability $P(\tau)$ for 
all variables $\tau_i$:
\begin{equation}
P_{eq} ({\cal C}) = \prod_{i=1}^{N-1} P(\tau_i).
\end{equation}
The probability $P(\tau)$ may be deduced from 
the evolution rates of the terminal reptons 
$p_a$ and $p_b$ independently of $p_c$. The 
particular configuration with all $\tau_i = 0$ 
leads to $p_a P(0) = p_b P(1)$. Thus, we deduce 
$P(1) = p_a/(p_a+p_b)$ and $P(0) = p_b/(p_a+p_b)$. 
The curvilinear length of the chain $L = 
\overline{s_N - s_1} = (N-1) a P(1)$ where the 
overline stands for an average over the 
configurations with the equilibrium probability 
$P_{eq}({\cal C})$. This result compares with the 
equilibrium length $N a + L$ in the model of Guidoni 
et al.~\cite{Guidoni} where the $N a$ difference 
comes from the length of the $N$ reptons. 

From Eq.~\ref{EqDif}, we deduce a differential 
equation for the average curvilinear position 
of the center of mass:
\begin{eqnarray}
\label{CurvPos}
\frac{\partial}{\partial t} \langle s \rangle ({\cal 
C},t) & = & - \sum_{i=0}^{N-1} w({\cal C} \rightarrow 
{\cal C}_i) \langle s \rangle ({\cal C},t) \\
\nonumber & + & \sum_{i=0}^{N-1} w({\cal C}_i 
\rightarrow {\cal C}) \left[\langle s \rangle 
({\cal C}_i,t) - \Delta s ({\cal C}) 
\right] \frac{P_{eq}({\cal C}_i)}{P_{eq} ({\cal C})}
\end{eqnarray}
where we used the property that $\int s^k P({\cal C},
s+\Delta s,t) ds = \langle (s-\Delta s)^k \rangle ({\cal 
C},t) P({\cal C},t)$ with $P({\cal C},t) = P_{eq}({\cal 
C})$. In the long time limit, the averaged curvilinear 
positions $\langle s \rangle ({\cal C},\infty)$ 
saturate and the left hand part of Eq.~\ref{CurvPos}
vanishes. The same line of arguments on the averaged 
curvilinear squared positions leads to the determination 
of the curvilinear diffusion coefficient $D_c(N)$. 
The long time limit of $\partial \overline{\langle s^2 
\rangle}/\partial t$ equals:
\begin{eqnarray}
\label{Curv}
\nonumber 2 D_c (N) & = & 2 \sum_{i=1}^{N-1} 
\overline{w ({\cal C} \rightarrow {\cal 
C}_{i}) \Delta s ({\cal C}) \langle s \rangle 
({\cal C},\infty)} \\
& + & \frac{a^2}{N^2} \sum_{i=1}^{N-1} \overline{w 
({\cal C} \rightarrow {\cal C}_{i})}.
\end{eqnarray}
After some transformations, the long time limit of
the squared positions $\langle s^2 \rangle ({\cal C},
\infty)$ cancelled out. Consequently, $D_c$ depends 
on the rates $w$ (Table~\ref{duke}) and the average 
curvilinear positions $\langle s \rangle ({\cal C},
\infty)$. For the later, we assume the following 
general expression:
\begin{equation}
\label{Express}
\langle s \rangle ({\cal C},\infty) = a 
\sum_{i=1}^{N-1} f(i) \, \tau_i
\end{equation}
where $f(i)$ is a function to be determined.
Consider the configuration with all variables $\tau_i 
= 0$, Eq.~\ref{CurvPos} leads to $f(1) = - f(N-1)$. 
For the configurations ${\tilde {\cal C}_k}$ 
with $\tau_i = 0$ except $\tau_k = 1$, if $2 
\leq i \leq N-2$, we obtain $f(i+1) = 2 f(i) 
- f(i-1)$ and, for $i=1$, $f(2) = (\Delta + 1) 
f(1) + (\Delta - 1)/N$ with $\Delta = (p_a + 
p_b)/p_c$. Note the particular case $\Delta = 
1$ or $p_a + p_b = p_c$ for which $f(i) = 0$ 
is a trivial solution. This particular case 
corresponds to the lack of correlations as 
discussed by Guidoni et al.~\cite{Guidoni}. 
In the case $\Delta \neq 1$, $f(i) = A(\Delta) \, 
i + B(\Delta)$ with $A(\Delta) = -2B(\Delta)/N$ 
and $B(\Delta) = (1-\Delta)/[(N-2) \Delta + 2]$. 
From this expression, it is possible to deduce 
$\langle s \rangle ({\cal C},\infty) = 2 B(\Delta) 
\delta s$. It is easy to check that the expression 
for $\langle s \rangle ({\cal C},\infty)$ given
by Eq.~\ref{Express} satisfies Eq.~\ref{CurvPos}
for all configurations ${\cal C}$. 

Two limiting cases are of interest. The fast 
internal motion of reptons compared to the 
end-repton motion corresponds to $\Delta 
\rightarrow \infty$. In this case, $\langle s 
\rangle ({\cal C},\infty) \simeq -2 \, \delta 
s/(N-2)$. In the case of slow internal motion 
of reptons, $\Delta \rightarrow 0$ and $\langle 
s \rangle ({\cal C},\infty) \simeq \delta s$. 
It is interesting to notice the change of sign 
in the prefactor between those two limiting 
cases implying a shift $\langle s \rangle 
({\cal C},\infty)$ in different directions
for the same difference $\delta s$ between the
center of mass and middle position of the chain. 

The exact curvilinear diffusion coefficient 
$D_c (N)$ is deduced from Eq.~\ref{Curv} and: 
\begin{eqnarray}
\sum_{i=0}^{N-1} \overline{w({\cal C} \rightarrow 
{\cal C}_i)} = 2 (N-2 + 2 \Delta) p_c P(0) P(1) \\
\sum_{i=0}^{N-1} w({\cal C} \rightarrow {\cal C}_i) 
(\tau_{i+1}-\tau_i) = p_c (\Delta - 1) (\tau_1 -
\tau_{N-1})\\
\overline{(\tau_1-\tau_{N-1}) \langle s \rangle 
({\cal C},\infty)} = 2 a P(0) P(1) f(1)
\end{eqnarray}
leading to:
\begin{equation}
D_c (N) = \frac{p_a p_b}{p_a + p_b} \, 
\frac{a^2}{(N-2) \Delta + 2} = \frac{D_0 
P(0) P(1)}{N - 2 + 2\Delta^{-1}}
\label{ExactCurv}
\end{equation}
with $D_0 = p_c a^2$ the individual diffusion 
coefficient of a free internal repton. This 
result is exact for all polymer lengths $N 
\geq 2$ and for all sets of rates $p_a$, $p_b$ 
and $p_c$. Note the linear dependence of 
$D_c^{-1}$ with the number of reptons:
\begin{equation}
D_c^{-1}(N) = \frac{N - 2 + 2 \Delta^{-1}}{D_0 
P(0) P(1)}. \label{Diffinv}
\end{equation}

\begin{figure}
\begin{center}
\includegraphics[width=8cm]{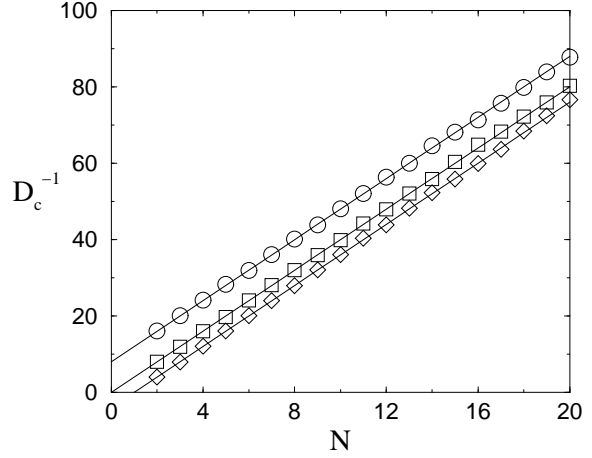}
\caption{\label{Dinverse}
Inverse of the curvilinear diffusion coefficient
$D_c$ as function of the number of reptons $N$ 
for different sets of rates. The lines correspond 
to the exact analytical calculations and the 
symbols to numerical simulations with $(p_a,p_b,p_c) 
= (1/4,1/4,1)$ circles, $(1/2,1/2,1)$ squares and 
$(1,1,1)$ diamonds.}
\end{center}
\end{figure}

Fig.\ref{Dinverse} illustrates this linear 
dependence for three different sets of parameters. 
The equality between the rates $p_a$ and $p_b$ 
implies identical equilibrium properties ($P(1) 
= P(0) = 1/2$). The individual diffusion coefficient
$D_0 = 1$ is imposed by setting $a = 1$ and $p_c = 1$ 
leading to an identical slope for all sets of rates 
considered. The sets only differ by the value of 
$\Delta$, with $\Delta = 1/2, 1$ and $2$ from top 
to bottom, leading to a shift of the different lines
on Fig.\ref{Dinverse}. The numerical results are
obtained from Monte-Carlo simulations. The average
of the curvilinear squared position as function of 
time for $10^5$ chains with initial conditions 
distributed with the equilibrium probability 
$P_{eq}({\cal C})$ are determined for different 
chain lengths. Linear fits allow us to determine 
the curvilinear diffusion coefficients. The 
numerical errors are smaller than the symbols 
on Fig.\ref{Dinverse}.

It is possible to understand Eq.~\ref{Diffinv} as 
follows. There exists $N-2$ equivalent internal 
reptons and $2$ extremal reptons. The inverse 
diffusion coefficient reflects the contribution 
of the two kinds of reptons. The free internal 
reptons have a diffusion coefficient $D_0$. The 
probability for them to move inside the chain is 
$P(0) P(1)$, the probability that there exists a 
unique stored length on the sides of the repton.  
This leads to the contribution $(N - 2)/D_0 P(0) 
P(1)$. The contribution of the two extremal 
reptons may be split in two parts: an extremal 
repton exploring a new cell (with a rate $p_a$ 
and a probability $P(0)$ to have a stored length 
on its side) and an extremal repton moving into 
the cell of its neighbor (with a rate $p_b$ and 
a probability $P(1)$ to have no stored length 
on its side). Both contributions are identical 
($p_a P(0) = p_b P(1)$) and add up to give the 
contribution $2 \Delta^{-1}/D_0 P(0) P(1)$ to 
the inverse diffusion coefficient $D_c^{-1}$ 
in Eq.~\ref{Diffinv}.

In a $d$-dimensional square lattice, the choice 
$p_a = p_c = 1$ and $p_b = 2d$ is customary and 
reflects the $2d$ possible directions for an 
extremal repton to explore a new cell compared 
to the single cell possibility for the other moves.
The curvilinear diffusion coefficient is then 
\begin{equation}
D_c(N) = \frac{2 d a^2}{(2d+1)[N(2d+1)-4d]}.
\end{equation} 
Exact results already obtained by Guidoni et 
al.~\cite{Guidoni} are recovered: the uncorrelated 
case $\Delta = 1$ leads to $D_c(N) = D_0 P(0) 
P(1)/N$ and the particular case $N = 3$ to 
$D_c(3) = D_0 P(0) P(1) \Delta/(\Delta + 2)$. 
However, as can be seen from Eq.~\ref{ExactCurv}, 
the second order in $1/N$ differs from the one 
proposed in~\cite{Guidoni} due to the improper 
account of the correlations in their calculations. 
For example, the limit $p_c \rightarrow 0$ or 
$\Delta \rightarrow \infty$ leads to the consistent 
limit of the curvilinear diffusion coefficient 
$D_c(N) \simeq a^2 P(0) P(1) p_c/(N-2) \rightarrow 
0$ in contrary to~\cite{Guidoni}. In contrary, 
the fast internal motion of reptons ($\Delta 
\rightarrow 0$ or $p_c \rightarrow \infty$) 
leads to a finite $D_c = P(0) P(1) a^2 /2$ independent 
of $N$.

\section{Periodic boundary conditions}

In the following, we consider the case of 
periodic boundary conditions studied by van 
Leeuwen and Kooiman~\cite{vanLeeuwen,Kooiman}. 
Those conditions correspond to introduce two 
new variables $\tau_N$ and $\tau_0$ with 
$\tau_N = \tau_0$. The rates $w({\cal C}_i 
\rightarrow {\cal C})$ and $w({\cal C} 
\rightarrow {\cal C}_i)$ for $i=0$ to $N-1$ 
have the same expression given in 
Table~\ref{duke} for $i=1$ to $N-2$. 
The number of independent variables $\tau_i$ 
in the periodic boundary case is increased to 
$N$ compared to the $N-1$ variables in the 
open boundary case leading to an increased 
number of configurations by a factor $2$. 

\begin{figure}
\begin{center}
\includegraphics[width=8cm]{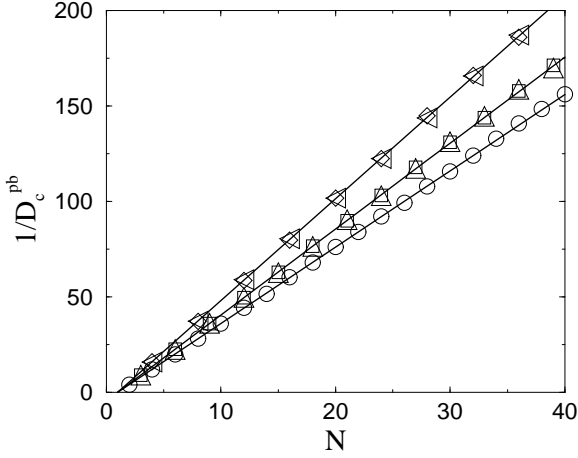}
\caption{\label{Dinversepb}
Inverse of the curvilinear diffusion coefficient 
$D_c^{pb}$ as function of the number of reptons $N$ 
for different probabilities $P(1)$. The lines 
correspond to the exact analytical calculations 
and the symbols to numerical simulations with 
$P(1) = 1/2$ circles, $1/3$ squares, $1/4$ 
diamonds, $2/3$ up triangles and $3/4$ left 
triangles.}
\end{center}
\end{figure}

In the case of periodic boundary conditions, 
the number $N_{1}$ of non-zero variables 
$\tau_i$ or $N_0 = N - N_1$ of stored lengths 
is conserved by the dynamics. In this respect, 
the dynamics is non-ergodic. All configurations 
${\cal C}$ with the same number of stored lengths 
have the same probability $P(N_0)$ and the 
equilibrium curvilinear length of the polymer 
with $N_0$ stored lengths is $L_{pb} = 
\overline{s_N - s_1} = (N-1) a \overline{\tau_i} 
= (N-1) a N_1/N$ since only $N-1$ variables 
$\tau_i$ are present in $s_N - s_1$. Furthermore,
\begin{eqnarray}
\sum_{i=0}^{N-1} \overline{w({\cal C} \rightarrow 
{\cal C}_i)} = 2 p_c N_1 (N - N_1)/(N - 1)\\
\sum_{i=0}^{N-1} w({\cal C} \rightarrow {\cal C}_i) 
(\tau_{i+1}-\tau_i) = p_c (\tau_N - \tau_0) = 0
\end{eqnarray}
lead to:
\begin{equation}
D_c^{pb} (N,N_1) = \frac{a^2 p_c N_1 (N-N_1)}{N^2 
(N-1)} = \frac{D_0 P(0) P(1)}{N-1}
\label{ExactCurvPB}
\end{equation}
where we replaced $N_0/N$ by the probability $P(0)$ 
to have a stored length and $N_1/N$ by $P(1) = 1 - 
P(0)$ in analogy with the open boundary case.
Note that the periodic and open boundary conditions 
present to the same leading behavior for the curvilinear 
diffusion coefficient $D_c \sim D_0 P(0) P(1)/N$ 
in the long chain limit. This result is also obtained 
for the self-diffusion coefficient but with a different 
length dependence~\cite{vanLeeuwen,Kooiman}.

Fig.\ref{Dinversepb} illustrates the linear 
dependence of the inverse curvilinear diffusion 
coefficient for different probabilities $P(1)$
as function of the number of reptons $N$. 
The numerical results are obtained from 
Monte-Carlo simulations by linear fits of the 
average of the curvilinear squared position as 
function of time for $10^5$ chains with initial 
conditions comprising $N_1 = N P(1)$ variables 
$\tau_i = 1$. The numerical errors are smaller 
than the symbols.

\section{Discussion and conclusion}

\begin{figure}
\begin{center}
\includegraphics[width=8cm]{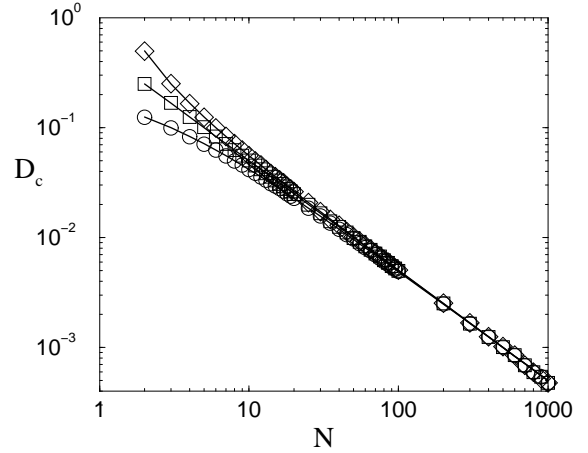}
\caption{\label{Dlog}
Curvilinear diffusion coefficient $D_c$ as function 
of the number of reptons $N$ in log-log scales
for different sets of rates. The lines correspond 
to the exact analytical calculations and the 
symbols to numerical simulations with $(p_a,p_b,p_c) 
= (1/4,1/4,1)$ circles, $(1/2,1/2,1)$ squares and 
$(1,1,1)$ diamonds.}
\end{center}
\end{figure}

In this paper, we have determined exactly the 
curvilinear diffusion coefficients $D_c (N)$ 
of the repton model. The cases of open and 
periodic boundary conditions were considered. 
The inverse curvilinear diffusion coefficients 
present a linear behavior with the number of 
reptons in both situations. The next to leading 
order term in the curvilinear diffusion coefficient 
$D_c$ shows interesting properties :
\begin{equation}
\frac{N D_c(N)}{D_0 P(0) P(1)} - 1 \simeq
\frac{2 (\Delta - 1)}{N \Delta}. 
\label{leading}
\end{equation}
This term is positive for $\Delta > 1$ and 
negative for $\Delta < 1$. The later case
would correspond to a chain with large 
end-reptons limiting their motion compared 
to the internal reptons. On Fig.\ref{Dlog},
$D_c(N)$ as function of $N$ on a log-log 
scale is presented for three different values
of $\Delta$, from bottom to top $\Delta = 
1/2,1$ and $2$. The change of concavity of 
the curves is representative of the sign 
change in Eq.~\ref{leading}. 

\begin{figure}
\begin{center}
\includegraphics[width=8cm]{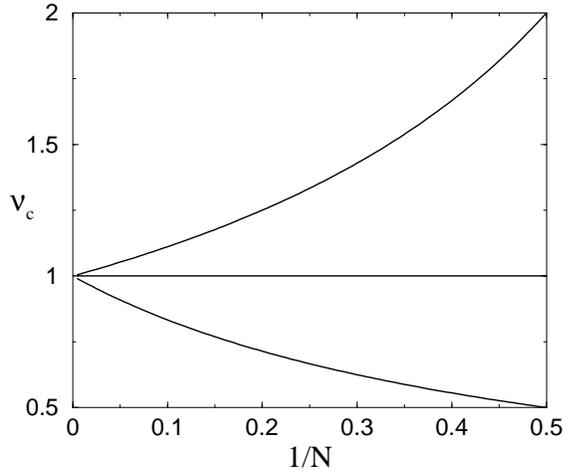}
\caption{\label{exponent}
Effective exponent $\nu_c$ as function of $1/N$ 
for three different values of $\Delta$. From 
bottom to top, $\Delta = 1/2,1$ and $2$.}
\end{center}
\end{figure}

A possible consequence of the next to leading 
order term concerns the determination of the 
effective exponent obtain by a linear fit of 
the diffusion coefficient in a log-log scale. 
Let us define a finite size effective exponent 
$\nu_{c}(N)$ for the curvilinear diffusion 
coefficient $D_{c}(N)$ as follows: 
\begin{equation}
\nu_c = \frac{\partial \ln D_c}{\partial \ln N} 
= 1 + \frac{2 (\Delta - 1)}{(N-2) \Delta + 2}
\end{equation}
Depending on $\Delta$, this exponent is either 
larger or smaller than the expected one $\nu_c 
= 1$ in the long chain limit except for the
particular case $\Delta = 1$ where $\nu_c(N) = 1$ 
(see Fig.\ref{exponent}). The non-monotonous behaviour 
observed for the self-diffusion $D_s$~\cite{Carlon} 
is not present for the curvilinear diffusion $D_c$. 
This difference is due to the absence of an anomalous 
behavior for $D_c$ whereas $D_s$ presents corrections
of order $1/N^{1/2}$ due to fluctuations in the chain 
length~\cite{Deutsch}.

\end{document}